\documentclass[notitlepage,twocolumn,pre,tightenlines,superscriptaddress,showpacs]{revtex4}
\usepackage{amsmath,amssymb,amsfonts,latexsym}
\usepackage[colorlinks=true, linkcolor=blue, citecolor=blue]{hyperref}
\usepackage{bbm}
\usepackage[mathcal]{euscript}
\usepackage{graphicx}
\usepackage{epsfig}
\usepackage{color}
\usepackage[activate=normal]{pdfcprot}  

\usepackage{psfrag}
\usepackage{wasysym}

\usepackage[caption = false]{subfig}

\usepackage{soul}  

\newcommand{\be}{\begin{equation}}
\newcommand{\ee}{\end{equation}}
\newcommand{\ben}{\begin{equation*}}
\newcommand{\een}{\end{equation*}}
\newcommand{\ba}{\begin{eqnarray}}
\newcommand{\ea}{\end{eqnarray}}
\newcommand{\ie}{\textit{i.e.} }

\newcommand{\avrge}[1]{\left\langle  #1 \right\rangle }

\newcommand{\phop}{p_{hop}}

\begin{document}

\title{Attractive versus truncated repulsive supercooled liquids: \\ The dynamics is encoded in the pair correlation function}

\newcommand{\eccm}{UMR Gulliver 7083 CNRS, ESPCI ParisTech,
PSL Research University, 10 rue Vauquelin, 75005 Paris, France}
\newcommand{\AoLRI}{comme ça :
LRI, AO team, Univ. Paris-Sud, Universit\'e Paris-Saclay, 91405 Orsay, France}
\newcommand{\DpaUPenn}{Department of Physics and Astronomy, University of Pennsylvania, Philadelphia, PA 19104}
\newcommand{\LPENS}{Laboratoire de Physique de l'\'Ecole normale sup\'erieure, ENS, Universit\'e PSL, CNRS, Sorbonne Universit\'e, Universit\'e Paris-Diderot, Sorbonne Paris Cit\'e, Paris, France}
\newcommand{\DChemColumbia}{Department of Chemistry, Columbia University, 3000 Broadway, New York, NY 10027, USA}

\author{Fran\c cois P. Landes}
\affiliation{LRI, AO team, B\^at 660 Universit\'e Paris Sud, Orsay 91405, France}
\email[Corresponding author: ]{francois.landes@u-psud.fr}
\affiliation{Laboratoire de Physique de l'\'Ecole normale sup\'erieure, ENS, Universit\'e PSL, CNRS, Sorbonne Universit\'e, Universit\'e Paris-Diderot, Sorbonne Paris Cit\'e, Paris, France}
\affiliation{Department of Physics and Astronomy, University of Pennsylvania, Philadelphia, PA 19104}
\affiliation{Department of Chemistry, Columbia University, 3000 Broadway, New York, NY 10027, USA}
\author{Giulio Biroli}
\affiliation{Laboratoire de Physique de l'\'Ecole normale sup\'erieure, ENS, Universit\'e PSL, CNRS, Sorbonne Universit\'e, Universit\'e Paris-Diderot, Sorbonne Paris Cit\'e, Paris, France}
\author{Olivier Dauchot}
\affiliation{UMR Gulliver 7083 CNRS, ESPCI ParisTech,
PSL Research University, 10 rue Vauquelin, 75005 Paris, France}
\author{Andrea J. Liu}
\affiliation{Department of Physics and Astronomy, University of Pennsylvania, Philadelphia, PA 19104}
\author{David R. Reichman}
\affiliation{Department of Chemistry, Columbia University, 3000 Broadway, New York, NY 10027, USA}

\begin{abstract}
We compare glassy dynamics in two liquids that differ in the form of their interaction potentials. Both systems have the same repulsive interactions but one has also an attractive part in the potential. These two systems exhibit very different dynamics despite having nearly identical pair correlation functions.
We demonstrate that a properly weighted integral of the pair correlation function, which amplifies the subtle differences between the two systems, correctly captures their dynamical differences.
The weights are obtained from a standard machine learning algorithm.
\end{abstract}

\maketitle

The central tenet of liquid state theory is that the structural and thermodynamic properties of liquids are primarily governed by the short-range repulsive part of their interaction potentials~\cite{widom_intermolecular_1967,Weeks1971}. The extension of these ideas to dynamical properties, however, is still in question.
Studies of glass-forming liquids with purely repulsive potentials show that their relaxation times $\tau_\alpha$ are well-described using a perturbative analysis around the hard-sphere limit~\cite{haxton,schmiedeberg}.
However, a study of the Kob-Andersen Lennard-Jones (LJ) mixture and its repulsive counterpart, the Weeks-Chandler-Andersen (WCA) mixture, found that the two systems have very different dynamics despite their structural similarities~\cite{berthier2009nonperturbative,Berthier2011}.
In particular, the relaxation time $\tau_\alpha$ is much smaller in the WCA system than in the LJ one at the same supercooled temperatures.
While hidden potential scale invariances \cite{pedersen_repulsive_2010}, complex structural indicators such as the point-to-set length  \cite{Biroli2004} and triplet correlation functions markedly differ in these two systems \cite{hocky_growing_2012,coslovich2013static},
the pair correlation functions are instead very similar. These findings indicate that long-ranged attractions strongly affect dynamics while leaving simple structural features relatively unaffected.

This result is a challenge to theories directly based on the exact (numerically computed) thermodynamically-averaged pair correlation functions $\avrge{g(r)}$ \cite{berthier2011testing}.
Mode coupling theory is the most prominent example of such a theory, and indeed it was shown to
largely underestimate the difference in dynamics of the WCA and LJ mixtures~\cite{Berthier2010}. From this negative result, it was argued that $\avrge{g(r)}$ does not contain the physical information relevant for predicting the dynamical slowdown, at least not in a way amenable to predictions, and that any theory of the glass transition based on the sole basis of $\avrge{g(r)}$ is bound to fail.

The view that $\avrge{g(r)}$ alone cannot explain the dynamical difference between the WCA and LJ systems has recently been questioned.
Combining it with the force-force correlation function and treating the attractive and slowly varying forces separately, Schweizer et. al. \cite{dell_microscopic_2015} proposed an alternative microscopic theory for the dynamical differences between LJ and WCA systems.
Modeling dynamics as that of a single-particle in an effective caging potential and using the pair-correlations based configurational entropy, \cite{nandi_role_2017} points toward $\avrge{g(r)}$ containing enough information to predict the transition temperature.
A comparison of that theory with Schweizer's has recently shown they produce comparable results, although the latter is more precise~\cite{saha_comparative_2019}.
Independently, approximate measures of the configurational entropy based on $\avrge{g(r)}$ can be used to estimate the dynamics via the Adam-Gibbs relation, and it was argued that the resulting difference of configurational entropy can explain the dynamical difference between the two systems ~\cite{Banerjee2014,Banerjee2016,Banerjee2017}. Although it is unclear why the configurational entropy should play a role at the temperatures studied, which are above the mode-coupling temperature, these results~\cite{Banerjee2014,Banerjee2016,Banerjee2017} suggest that the relationship between $\avrge{g(r)}$ and dynamics should be revisited.

In this RRapid Communication we demonstrate that a properly weighted integral of the pair correlation function obtained by machine-learning (ML) techniques~\cite{Cubuk2015,Schoenholz2016,Cubuk2016,Schoenholz2016a,Definiujemy2012,Sussman2016}, the so-called "softness" designed to correlate strongly with instantaneous mobility, fully captures the dynamical differences between the LJ and the WCA supercooled liquids, despite their minute differences in average structure. Our results offer a simple explanation of the relationship between
$g(r)$
and relaxation time, and show that softness is an important structural fingerprint of dynamics.
We emphasize that in contrast to previous works on Softness, here we are able to extract dynamics in systems with very similar $\avrge{g(r)}$, showing that fluctuations in $g(r)$ are relevant to the dynamics.
We conclude that, at least in principle, one should be able to compute the relaxation time of these two liquids on the sole basis of the pair correlation.

\begin{figure}  [t!]
\centering
\includegraphics[width=\columnwidth]{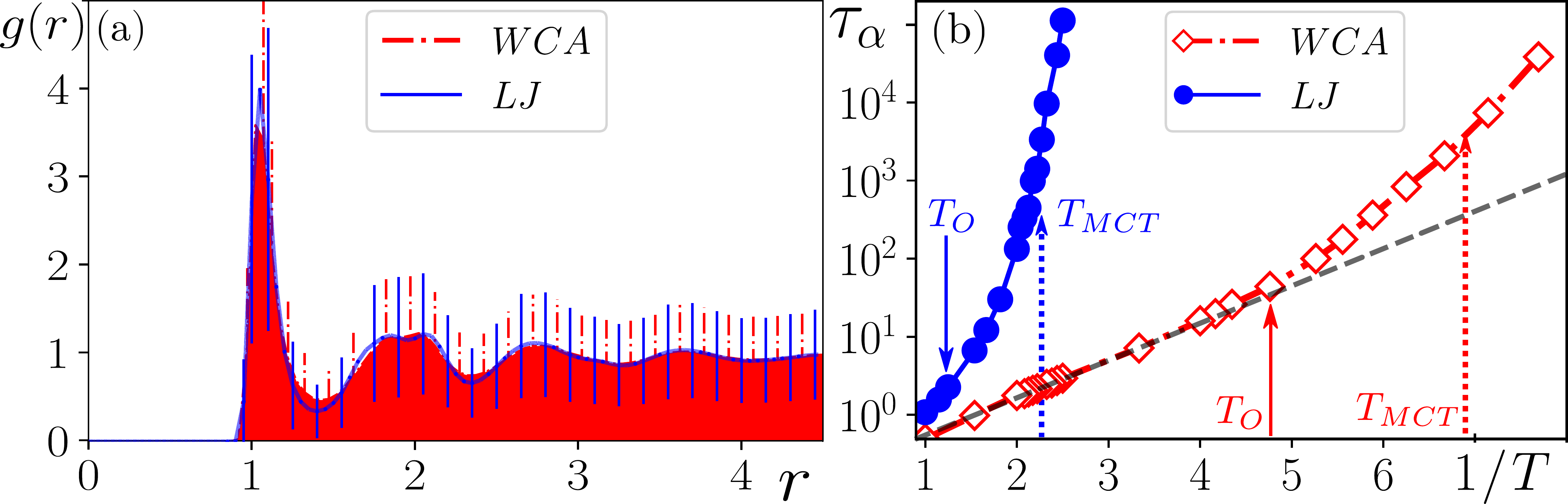}
\caption{{\bf Structure and dynamics for WCA and LJ systems.}
\textbf{(a)} The $AA$ pair correlation function $\avrge{g_{AA}(r)}$ at $T=0.43$, for the LJ liquid (blue), and the WCA one (red); (vertical bars denote the standard deviation of $P_r(g)$).
\textbf{(b)} Temperature dependence of the relaxation time $\tau_{\alpha}(1/T)$ for the two systems.
Arrows indicate the onset and MCT temperatures.
The black dashed line outlines the initial Arrhenius behavior of the WCA liquid.}
\label{fig:conun}
\vspace{-3mm}
\end{figure}

Following Berthier and Tarjus~\cite{Berthier2011}, we compare two Kob-Andersen~\cite{Kob1995,Kob1995a} binary mixtures of A and B particles, one with Lennard-Jones interactions and the other with WCA interactions~\cite{Weeks1971}.
The latter system shares the same repulsive short-range interaction with the LJ system, but possesses no attractive forces, \ie the potential is zero at the minimum $r=r_{cut}=2^{1/6} \sigma_{XY} $ and beyond. The potentials are smoothed at the cutoff distance, so that their second derivatives are continuous everywhere.
The two liquids are first equilibrated in the NVT ensemble; we then record particle trajectories $\vec{r}_i(t)$ for $\sim 200 \tau_\alpha$ in the NVE ensemble \cite{
anderson_general_2008,glaser_strong_2015} and extract the relaxation time $\tau_\alpha$ from the decay of the self-intermediate scattering function.
Following Ref.~\cite{Berthier2011}, we show in Fig.~\ref{fig:conun}b that $\tau_\alpha$ increases much more rapidly in the LJ system than in the WCA one when temperature is lowered, while concomitantly the differences in the average pair correlation function $\avrge{g(r)}_{AA}$ between systems remain small relative to the fluctuations within each system, as shown in Fig.~\ref{fig:conun}a.

In the supercooled regime particles essentially fluctuate around a given average position, effectively trapped in a {\it cage} of surrounding particles.
From time to time a rearrangement, involving a small number of particles, occurs.
At the level of a single particle $i$, a rearrangement is characterized by a {\it cage-jump}, corresponding to a relatively rapid change in the average position of the particle that is on the scale of the particle size.
Cage-jumps taking place during a time interval $W=[t_1,t_2]$, for a given particle $i$, may be captured by computing the quantity $p_{hop}(i,t)$~\cite{Candelier2009,Candelier2010a,Candelier2010c,Smessaert2013a}:
\begin{align}
p_{hop}(i,t)
 \equiv
\sqrt{ \avrge{ \left( \vec{r}_i - \avrge{\vec{r}_i }_U \right) ^2 }_V ^{1/2}   \avrge{\left(\vec{r}_i - \avrge{\vec{r}_i }_V\right)^2 }_U  ^{1/2} }\,\, ,
\end{align}
for all $t \in W$, where averages are performed over the time intervals surrounding time $t$, i.e. $U=[t_1,t]$, $V=[t,t_2]$.
$t^* = \text{argmax}_{t\in W} (\phop(i,t))$ is the time that best separates the trajectory $\vec{r}_i([t_1,t_2])$ into two sub-trajectories  $\vec{r}_i([t_1,t^*])$ and $\vec{r}_i([t^*,t_2])$. We record $\phop^*(i)= \phop(i,t^*)$ and the process is iterated on each sub-trajectory. If the trajectory during the time interval $W$ is contained within a cage then $\phop^*(i)$ is small and corresponds to the cage's size.
Conversely, if $W$ contains two distinct cages then $\phop^*$ is large and corresponds to the jump's amplitude.
We thus interrupt the iteration when $\phop^* < p_c$, where $p_c$ is chosen as the root-mean-squared displacement (RMSD) $\Delta r(t)$, computed at the time where the non-Gaussian parameter $\alpha_2 = \frac{3 \avrge{r^4}(t) }{5\avrge{r^2}^2(t) } - 1$ has a maximum. This choice is rather stringent: only displacements several times larger than the plateau of the RMSD are considered as jumps.
Finally, as jumps are never instantaneous, we assign $\phop(i,t) = \phop^*(i)$ to all times $t\in [t^*-t_f, t^*+t_f]$, where $t_f=5$ in LJ-time units, i.e. about five ballistic times. This procedure generates a label that classifies every particle at every instant as
caged or jumping. This constitutes what is called in ML jargon the ground truth: $y_{GT}(i,t) \equiv \Theta(\phop(i,t)-p_c)\in \{0,1\}$.

If dynamics is indeed related to local structure, then there should be small but significant structural differences between the local neighborhoods of caged and cage-jumping particles, respectively. A number of approaches have attempted to identify these subtle structural signatures in supercooled liquids~\cite{Coslovich2007,chen_low-frequency_2010, zylberg_local_2017}.
They all have in common the fact that they are based on a deep knowledge of the physics of liquids.
Here, in contrast, we use a ML approach as introduced in Refs.~\cite{Cubuk2015,Schoenholz2016,Schoenholz2016a}.
The first step is to identify the structural observables that we will try to correlate with dynamics. We focus on
$\vec{G}(i,t)$, a vector of $160$ \textit{descriptors} of the local structure around particle $i$ at time $t$. Considering particles of type $A$, it reads:
\begin{align}
G_{Y,\alpha}(i,t)= \frac{1}{4 \pi r_\alpha^2} \cdot
\sum_{j\in \{Y\}} \mathbb{I}_{ |\vec{r}_{j}-\vec{r}_{i}|\in[r_\alpha, r_{\alpha+1}] },
\label{eq:descriptors}
\end{align}
where $Y$ denotes the type of particles $(A,B)$, and $r_\alpha~=~\{0.70,0.75,...,4.70\}$ is an $80$-point binning of the distance separating pairs of particles.
Note that $G_{Y,\alpha}(i,t)$ and the instantaneous pair correlation, $g_{X_iY}(r_\alpha,t)$, are proportional up to a constant that depends on the binning size.
\begin{figure}[t!]
\centering
\includegraphics[width=\columnwidth]{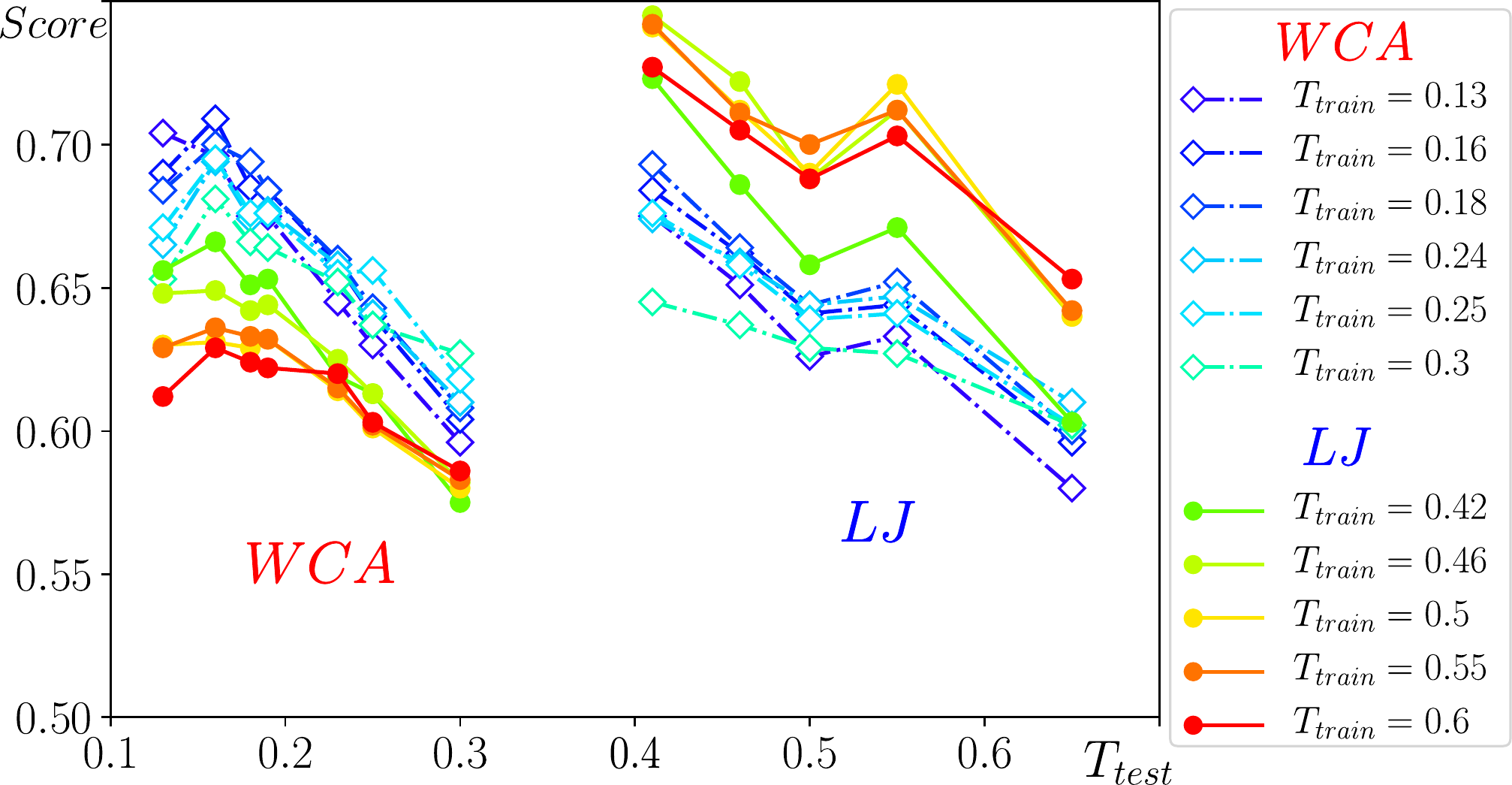}
\caption{
\textbf{Structure-dynamics relationships.}
Learning scores obtained using different training sets (WCA or LJ, $T_{train}$, as indicated in the legend) and applied to test sets (WCA left hand side, LJ right hand side), at all available temperatures $T_{test}$.}
\label{fig:cross-table-scores}
\vspace{-3mm}
\end{figure}
The second step, called training in the context of ML, consists in fitting a function $f(\vec{G}(i,t)), sign(f(G(i,t)))=y_{GT}(i,t)$, that allows to predict unobserved occurences of $y_{GT}$ from the corresponding $\vec{G}(i,t)$. 
To do so, we use a linear support vector machine~\cite{guyon_automatic_nodate,scikit-learn}, i.e. the function $f$ is an affine function of the input vector's coefficients. An intuitive picture is to consider $f$ as encoding the hyperplane which best separates the two populations $y_{GT}(i,t) = -1$ or $1$, in the 160-dimensional descriptor space. In our context, the generic fitting function $f$ is called softness $S$:
\begin{align}
S(i,t) = \vec{w} \cdot \vec{G}(i,t) + b,
\label{eq:softness}
\end{align}
namely the signed distance between the point $\vec{G}(i,t)$ and the hyperplane defined by $b$ and $\vec{w}$. 
Concretely, for a given training set, an iterative optimization routine adapts the coefficients $\vec{w}, b$ so as to minimize the number of errors on the training examples (an error is when $ sign(f(G(i,t)) \neq y_{GT}(i,t)$.
One says the model generalizes well when the error rate on a test set is as low as the training error rate (the test set it typically drawn from the same distribution as the training set, but completely uncorrelated from it).
The softness characterizes the local, instantaneous structure of any particle for the two systems (LJ, WCA), at any temperature.
Note that $\vec{G}(i,t)$ and $\phop(i,t)$ are computed using the actual finite-temperature structures, similar to the Method of \cite{Sussman2016}; unlike that in \cite{Schoenholz2016a,Schoenholz2016}, we do not use inherent structures (quenches to the local $T=0$ configuration).
The training set is built by sampling uniformly at random $N_{train}=10^4$ spatio-temporal instants $(i,t)$ with  $\phop^*(i,t)>p_c$, and the same number from the other class.
Finally, the third step of the procedure tests whether the algorithm classifies the data correctly by checking the performance on data that have not been used for training.
For any choice of training set (WCA or LJ, $T_{train}$), one can apply the hyperplane $\vec{w}$ to all other test data (WCA or LJ, $T_{test}$) and measure the average classification accuracy (rate of correct answers) on balanced test sets ($50/50$ mix of $y_{GT}=-1$ and $y_{GT}=1$):
\begin{align}
\text{Score} = \frac{\mathbb{P}(S>0|y_{GT}=1) + \mathbb{P}(S<0|y_{GT}=1) }{2},
\end{align}
where balance is obtained by downsampling until $\mathbb{P}(y_{GT}=-1)=1/2=\mathbb{P}(y_{GT}=1)$.
Fig.~\ref{fig:cross-table-scores} shows how this score depends on the choice of training set for various test sets.
As expected the score decreases with increasing $T_{test}$: hotter is harder to predict because mobility is less correlated with structure. More interesting is that $T_{train}$ is not crucial:  curves corresponding to different $T_{train}$ largely superimpose for the same training system.
The choice of training system, however, is important: hyperplanes trained on a given system perform significantly better on that system.
Overall, the high scores we achieve demonstrate a strong structure $-\vec{G}(i,t)-$ to mobility $-\phop(i,t)-$ relationship
encoded in the softness.

\begin{figure}[t!]
\centering
\includegraphics[width=\columnwidth]{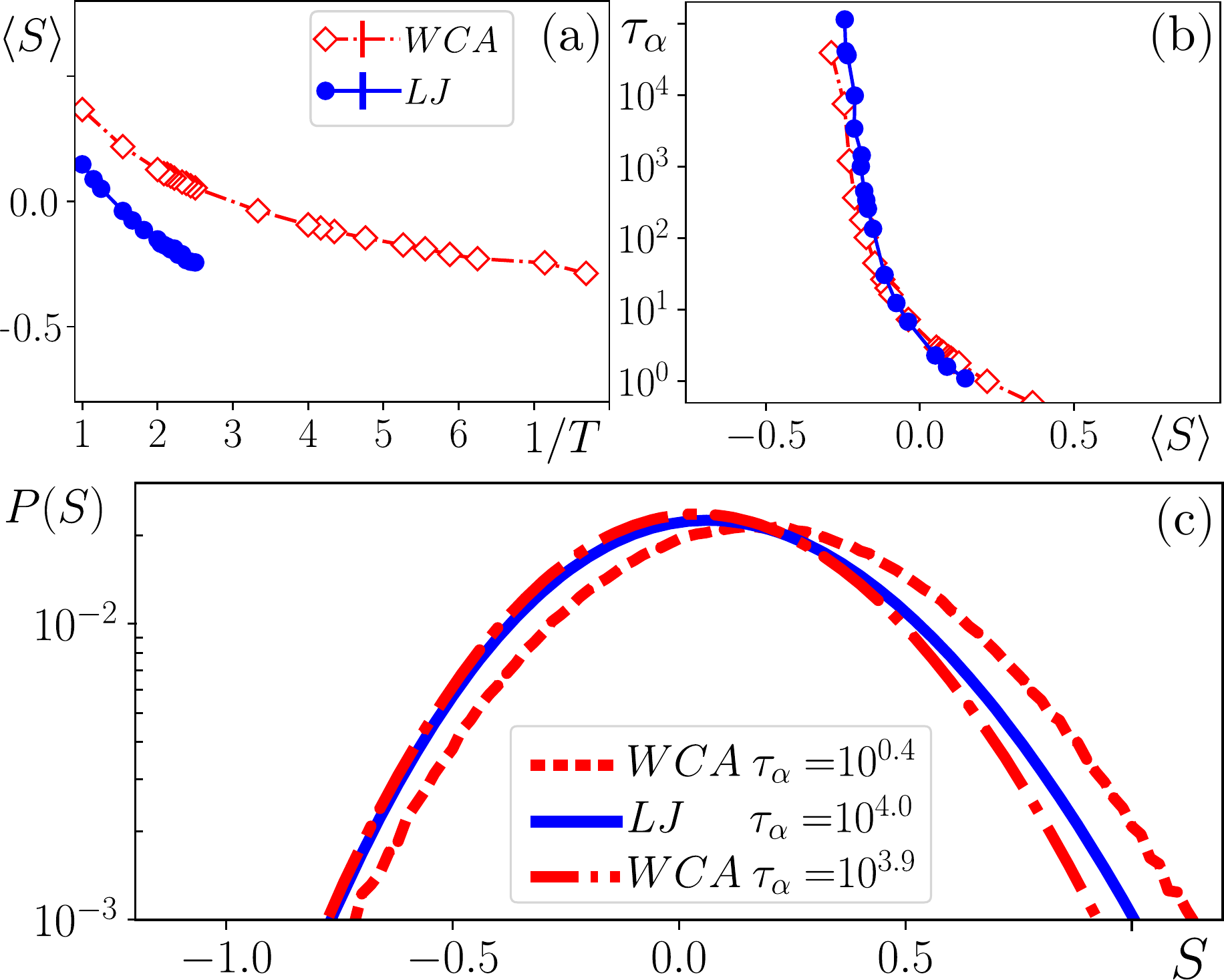}
\caption{\textbf{Dynamics and Softness.}
\textbf{(a)} Average softness $\avrge{S}$ as a function of inverse temperature and
\textbf{(b)} relaxation time, $\tau_\alpha$, as a function of $\avrge{S}$  for for LJ and WCA.
The span of the $\avrge{S}$ axis respects the dispersion of $P(S)$, as shown in panel (c).
\label{fig:softness}
\textbf{(c)} Probability distribution of the softness $\mathbb{P}(S)$ for the WCA and LJ systems at the same temperature $T=0.43$ (dashed and full line) and such that they share the same relaxation time $\tau_\alpha\approx10^4$ (full and dash-dotted line, the latter being the WCA system at $T=0.14$).
\label{fig:PofS}
}
\end{figure}

From now on, we shall set a unique hyperplane obtained by training at $T=0.43$, and using a balanced mix of WCA and LJ data. Other choices lead to similar (but slightly less good) results.  We compute the local and instantaneous softness for each liquid using Eq.~(\ref{eq:softness}) and average it over all trajectories. Fig.~\ref{fig:softness}(a) shows that the average softness $\avrge{S}(1/T)$, a purely structural quantity, is quantitatively different for the two systems. However, when plotting $\tau_\alpha$ as a function of $\avrge{S}$, the differences between the dynamics essentially disappear as shown in  Fig.~\ref{fig:softness}(b): not only does $\avrge{S}$ amplify minute differences in the structure, it does so in such a way that the softness, on average, is an excellent structural indicator of the dynamics.
Since the softness is linearly related to the structural descriptors, one simply has
\begin{align}
\avrge{S} = \vec{w} \cdot \avrge{\vec{G}} + b ,
\label{eq:avgsoft}
\end{align}
where $\vec{\avrge{G}} \approx \avrge{g}(r)$, in the limit of vanishingly small binning width.
We thus conclude that the difference in dynamics between the LJ and the WCA systems is, to a large extent, encoded in $\avrge{g}(r)$, but in a subtle way. The two populations of mobile and immobile particles are characterized by different instantaneous {\em local} $g(r)$, which our method allows to collapse into a scalar variable, the softness. Its
average value is therefore related to the average mobility and, hence, to the relaxation time.

Fig.~\ref{fig:softness}(c) shows that the width of the distribution of softness values is broad compared to the shift of $\langle S \rangle$ with temperature that is shown in Fig.~\ref{fig:softness}(a).
Fig.~\ref{fig:PofS} compares the distribution $P(S)$ obtained for the LJ and WCA systems at low temperature ($T=0.43$), and that for the WCA system at a different temperature, $(T=0.14)$, such that they share the relaxation time $\tau_\alpha\simeq 10^4$. As expected, the distribution for the WCA system is shifted towards higher softness values as compared to the LJ ones at the same temperature. However the distributions have a strong overlap, suggesting that the softness field is highly fluctuating. One should thus not expect a deterministic relationship between the local, instantaneous value of the softness of a particle and the local mobility.

The distributions for the WCA and LJ systems, compared at different temperatures but at the same relaxation time, are quite similar and share a common average.
Still, a systematic difference persists on the side of large softness, indicating again system-specific structure-mobility relationships. On the other hand, the negative-$S$ tails are very much alike. This last observation suggests that non-mobile, negative-$S$ particles play a system-independent role in controlling $\tau_\alpha$.
\begin{figure}[t!]
\centering
\includegraphics[width=0.99\columnwidth]{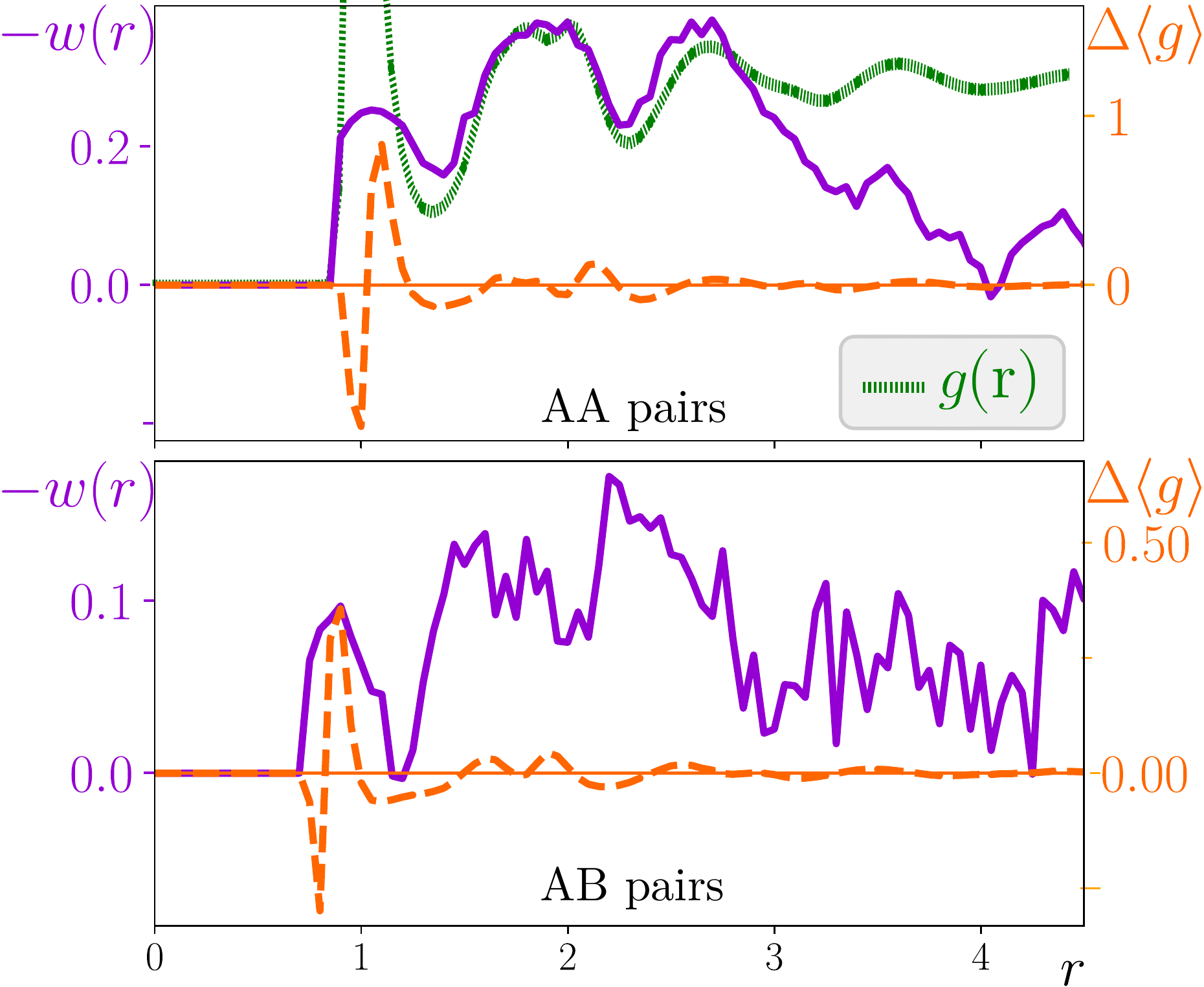}
\caption{{\bf Weighting the pair correlation function.}
    Weights $w_{AY}(r_\alpha)$, together with the difference $\Delta \avrge{g}=\avrge{g}_{WCA} - \avrge{g}_{LJ}$ for the A-A and A-B pairs of particles ($T=0.43$), along with the average $\avrge{g(r)}_{LJ}$, shown for comparison (in green).}
\label{fig:weights}
\vspace{-3mm}
\end{figure}

From a theoretical viewpoint, the weights $w$ in Eq.~\ref{eq:softness}, or equivalently the softness field, provide a promising route towards a better understanding of the relation between the structure and dynamics of supercooled liquids.
These weights encode the properties of the structural fluctuations, and more precisely the fluctuations of $g_{XY}(r)$. Moreover,
Eq.~(\ref{eq:avgsoft}) shows that the average softness $\avrge{S}$ is a weighted integral of the average pair correlation function, where the weights are the components of $\vec{w}$, obtained from the training phase in the ML process, which appears directly related to the relaxation time (Fig.~\ref{fig:softness}(b)).
Fig.~\ref{fig:weights} displays these weights $w_{AX}(r_\alpha)$, together with  $\Delta \avrge{g_{AX}}=\avrge{g_{AX}}_{WCA} - \avrge{g_{AX}}_{LJ}$, the minute differences between the average pair correlation functions for $X=A$ and $B$. High $w$ correspond to features most relevant to the dynamics.
Apart from the general observation that the short distance structure is crucial in triggering dynamics, the weighting is non-trivial; it is not a mere amplification of modulations in the pair correlation function.
The first minima and maxima of $\Delta \avrge{g_{AX}}$ are amplified, but less than the tiny differences observed in the structure of the second shell. Conversely the second minimum in $\Delta \avrge{g_{AY}}$  is strongly suppressed by the weighting.

\begin{figure}[t!]
\centering
\includegraphics[width=0.99\columnwidth]{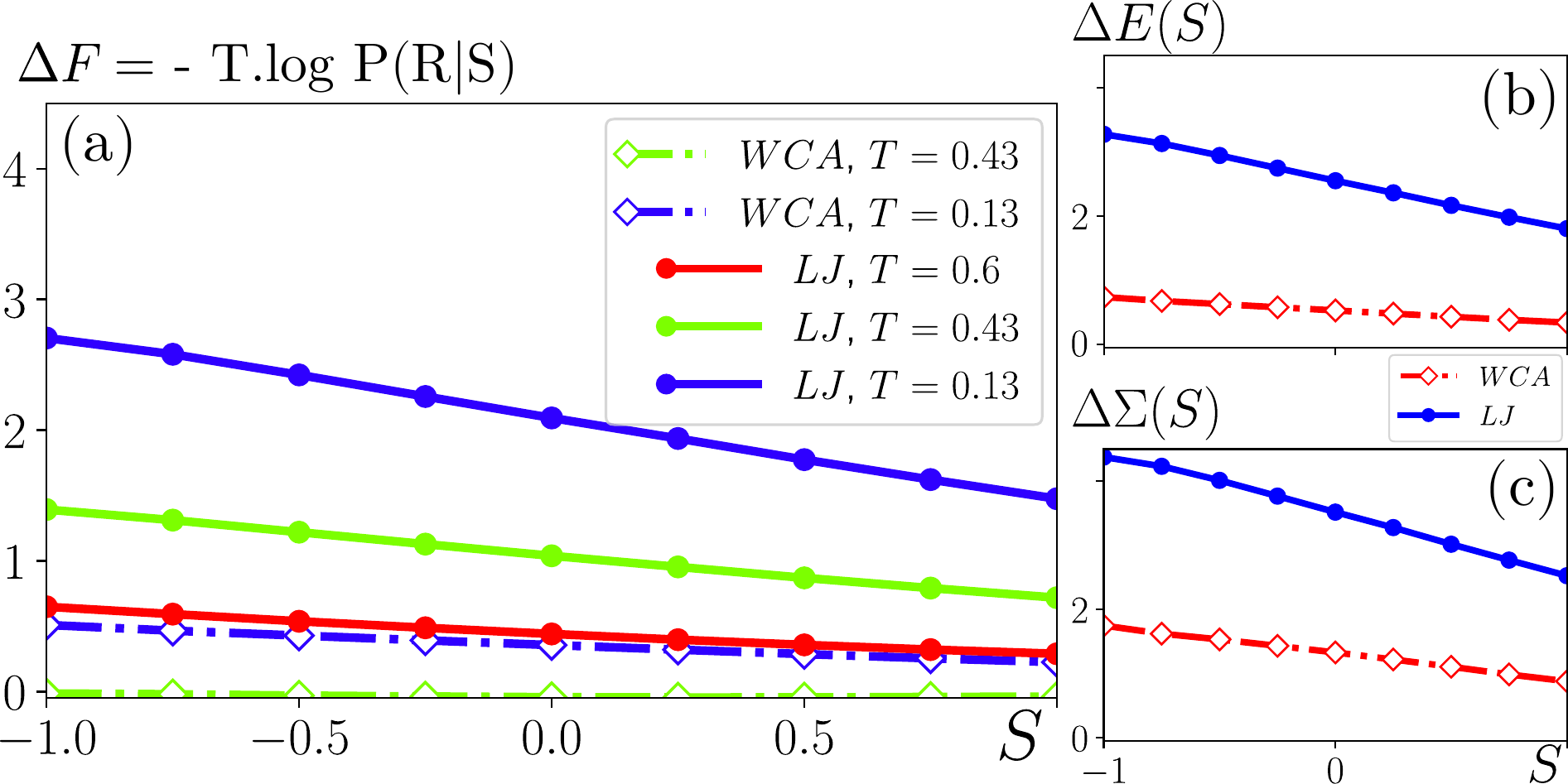}
\caption{
{\bf Energy and Entropy barriers.}
\textbf{(a)} The free energy barrier dependence in $S$, $\Delta F(S)$, directly related to the rate of activation $P(R|S)$.
\textbf{(b,c)} The coefficients $\Delta E(S), \Delta \Sigma(S)$ as defined in  Eq.~(\ref{eq:PRS}).
\label{fig:DeltaF}
}
\end{figure}

From the machine-learning point of view, we proceed in an unusual way since we trained on
a binary classification task (mobile, not mobile) but used the result as in a regression problem, i.e. focusing not only
on the sign of $S$ but also on its value. In order to determine whether the magnitude of $S$ has physical meaning, we follow ~\cite{Schoenholz2016,Schoenholz2016a} and compute $P(R|S)$ (left panel of Fig.~\ref{fig:DeltaF}, the probability of observing a local re-arrangement $(P(\phop(i,t)>\phop^*))$ for a given range of softness $\in [S, S+0.025]$, for different $1/T$.
As expected intuitively, large $S$ lead to a tighter
relationship with mobility, thus justifying the use of $S$ and not just its sign. As in~\cite{Schoenholz2016,Schoenholz2016a}
we find that $\log P(R|S)$ decreases linearly with $1/T$ and that
\begin {align}
P(R|S) = \exp \left[ - \frac{\Delta E(S) - T \Delta \Sigma(S)}{T} \right]
\label{eq:PRS}
\end{align}
Note that $\Delta E$ and $\Delta \Sigma$ depend far more strongly on $S$ for LJ than for WCA,
as shown in Fig.~\ref{fig:DeltaF}(b,c).
Following arguments of Ref.~\cite{Schoenholz2016}, the observed $S$ dependence leads to a significantly higher value of the onset temperature, $T_0$, for LJ than for WCA, consistent with prior calculations~\cite{Berthier2011}, as well as a stronger $T$-dependence as in Fig.~\ref{fig:conun}(b).

In summary, we have shown that the softness field provides a quantitative, structural characterization of the dynamical slowdown in the mode-coupling regime, which is precise enough to distinguish liquids with very similar structures, namely LJ and WCA mixtures of equal composition.
From the theoretical viewpoint many questions remain open with respect to the nature of the softness field.
Perhaps the most important of these questions is whether the kind of fluctuations picked up by the softness are those envisioned
in dynamical facilitation theory \cite{keys2011excitations}, connected to mobile defects, or instead are related to the self-induced disorder that facilitates motion within thermodynamic theories such as random first-order transition theory \cite{bhattacharyya2008facilitation,franz2011field,biroli2018random}. This question can be answered
by studying the dynamical evolution of the softness field, and by comparing it to the avalanche dynamics of
supercooled liquids. We will address this question in future work.

\begin{acknowledgments}
We thank S. S. Schoenholz and E. D. Cubuk for crucial input at the start of this project, D. Coslovich for helpful discussions, and L. Berthier and G. Tarjus for useful feedback on the manuscript 
This work was funded by the Simons Collaboration "Cracking the glass problem" via 
348126 (SRN), 454935 (GB), 327939 and 454945 (AJL),
and 454951 (DR).
\end{acknowledgments}

\end{document}